\documentclass[useAMS,usenatbib,12pt]{mn2e}
\usepackage[dvips]{graphicx,color}
\usepackage{epsfig}
\usepackage[latin1]{inputenc}
\usepackage{graphics}
\usepackage{amsfonts}
\usepackage{amsmath}
\usepackage{multicol}
\usepackage{layout}
\usepackage{amssymb}
\usepackage[a4paper]{hyperref}
\def\msun{{\rm\,M_\odot}}

\def\sr{{\rm\,sr}}

\def\kpc{{\rm\,kpc}}

\def\sr{{\rm\,sr}}

\def\cms{$\rm cm^3\;s^{-1}$}

\def\phcms{$\rm ph\;cm^{-2}\;s^{-1}$}
\def\B0{B$_{ref,0}$} 

\newcommand{\beq}{\begin{equation}}
\newcommand{\eeq}{\end{equation}}
\newcommand{\be}{\begin{equation}}
\newcommand{\ee}{\end{equation}}

\long\def\comment#1{}

\def\msun{M_{\odot}}
\newcommand{\ba}{\begin{eqnarray}}
\newcommand{\ea}{\end{eqnarray}}

\def\rsun{{R_\odot}} 

\title[Subhaloes in  a tree with microsolar  mass resolution]{A Merger
  Tree  with Microsolar Mass  Resolution: Application  to $\gamma$-ray
  Emission from Subhalo Population}

\author[C.   Giocoli,  L.   Pieri,  G.  Tormen  \&  J.   Moreno]{Carlo
  Giocoli$^{1,2,3}$, Lidia  Pieri$^{1,4,5}$, Giuseppe Tormen  $^1$ and
  Jorge             Moreno             $^6$             \thanks{Email:
    \href{mailto:cgiocoli@ita.uni-heidelberg.de}{cgiocoli@ita-uni-heidelberg.de},
    \href{mailto:pieri@iap.fr}{pieri@iap.fr},
    \href{mailto:giuseppe.tormen@unipd.it}{giuseppe.tormen@unipd.it},
    \href{mailto:jmoreno@haverford.edu}{jmoreno@haverford.edu}.}  \\
  $^1$ Dipartimento di Astronomia, Universit\`a degli Studi di Padova,
  Vicolo dell'osservatorio  2 I-35122  Padova, Italy \\  $^2$ Istituto
  Nazionale  di  Astrofisica  -  Osservatorio Astronomico  di  Padova,
  Vicolo  dell'osservatorio 2  I-35122 Padova,  Italy \\  $^3$ Zentrum
  f$\ddot{u}$r   Astronomie,  ITA,   Universit$\ddot{a}$t  Heidelberg,
  Albert-Ueberle-Str.  2, 69120  Heidelberg, Germany \\ $^4$ Consorzio
  Interuniversitario di Fisica Spaziale, Villa Gualino, Viale Settimio
  Severo,  63, I-10133  Torino, Italy  \\ $^5$  Istituto  Nazionale di
  Fisica Nucleare -  Sezione di Padova, Via Marzolo  8 I-35131 Padova,
  Italy \\ $^6$ Department of Physics \& Astronomy, Haverford College,
  370 Lancaster Avenue, Haverford, PA 19041, USA}

\begin{document}
\date{}
\pagerange{\pageref{firstpage}--\pageref{lastpage}} \pubyear{2007}
\maketitle
\label{firstpage}

\begin{abstract}
  The hierarchical growth of dark matter haloes, in which galaxies are
  hosted, has  been studied and modeled using  various approaches.  In
  this paper we  use a modified version the  Sheth \& Lemson algorithm
  for a $\mathrm{\Lambda}$ cold  dark matter power spectrum, and model
  the  growth   of  a  Milky-Way  sized  halo   with  microsolar  mass
  resolution,  corresponding to  the  typical Jeans  mass  for a  dark
  matter Weakly Interacting Massive Particle  with mass of 100 GeV. We
  then compute the \emph{unevolved} subhalo mass function and build-up
  a Milky-Way halo placing  and evolving its satellites.  This subhalo
  population  is used  to study  the $\gamma$-ray  emission  from dark
  matter annihilation.  In this case, the subhaloes which populate the
  host  halo have  been  computed considering  only progenitor  haloes
  accreted by the main branch of  the tree, so as to correctly treat the embedding
  of sub-subhaloes inside subhaloes.  Each subhalo will indeed host at
  the  present-time sub-subhaloes  accreted  when it  was an  isolated
  system.   In  order  to  compute  the sub-subhalo  population  of  a
  Milky-Way dwarf  galaxy, like Draco,  and to study  its $\gamma$-ray
  emission,  we  first  estimate  the  Draco virial  mass  at  merging
  redshift $z_m$ and then we  run the merger tree from $z_m$ following
  the halo  down to  the dark  matter Jeans mass.   We then  study the
  effect on the Fermi-LAT  (GLAST) detectability for both subhaloes in
  the  Milky-Way and in  Draco, and  we show  how subhaloes  cannot be
  responsible for the boost factor needed for detection.
\end{abstract}

  
\section{Introduction}

In  the standard scenario  of structure  formation galaxies  reside in
massive Dark  Matter (DM)  haloes, where baryons  can shock,  cool and
eventually  form  stars.  Haloes  form  by gravitational  instability,
starting  from  some   density  fluctuation  field  $\delta(\vec{x})$.
Specifically, a dark matter halo with mass $M \propto R^3$, forms when
the linear density field, smoothed on scale $R$, exceeds some critical
threshold $\delta_{c}$ \citep{bondetal91,carr,lc93,eke}.

In a Cold Dark Matter (CDM) framework, the galaxy formation process is
hierarchical along  cosmic time: small systems collapse  earlier, in a
denser  universe,   and  later  merge  to  form   larger  haloes.   In
particular,  haloes often grow  as a  consequence of  repeated merging
events with smaller  satellites. Present-day surviving satellites form
the  so-called subhalo (or  substructure) population  of a  given host
system \citep{moore,swtk01,gao,vtg05,gtv,zen}.

Dark  matter haloes  and subhaloes  provide the  environment  in which
galaxies form  and evolve \citep{bullapj,som02,krav,vale}.  Therefore,
understanding  the assembly  histories of  dark matter  haloes  is the
first  step towards the  comprehension of  the more  complex processes
involved in galaxy formation.

In order to study the  framework of structure formation, two different
approaches  have commonly  been used.   The first  is to  run $N$-body
simulations \citep{gadget,millennium,vialactea,dkm07}.  These are very
powerful tools,  that can  be used to  reproduce the collapse  of dark
matter haloes with high mass  and force resolution, both on galaxy and
galaxy-cluster scales. They allow  to follow gravitational collapse up
to its fully non-linear evolution.  On the other hand, simulations are
computationally  expensive,  and  nonetheless  cannot cover  the  full
spectrum  of  masses  relevant  to structure  formation.   The  second
approach is to  use some analytical modeling, which  allows a detailed
study of the merging history  of haloes over an arbitrarily large mass
range, under suitable simplification of the problem.  This is the case
of the Press \& Schechter formalism \citep{lc93,sheth98,sl99,sheth03}.

As  underlined before,  haloes collapse  on a  certain scale  once the
linear density contrast smoothed  on that scale exceeds some threshold
value. The  nonlinearities introduced  by these virialized  objects do
not affect the  collapse of overdense regions on  larger scales.  This
simple assumption leads to the  derivation of the global mass function
of dark  matter haloes that  is in fair  agreement with that  found in
$N$-body  simulations \citep{lc93,som00,st99,st02}.   During  the last
decades,  an extension  of this  theory  was made  by several  authors
\citep{lc93,bondetal91,bower91}   with  the   aim  of   computing  the
probability that  a halo of  mass $m$, at  redshift $z$, belongs  to a
given halo  of mass $M_0$ at redshift  $z_0<z$ \citep{lc93,st02}. This
quantity  was named ``conditional''  or ``progenitor''  mass function,
$f(m,z|M_0,z_0)\mathrm{d}m$.    This  allowed  different   authors  to
estimate quantities such as  the merger \citep{nd08} and creation rate
\citep{pm99,pmp00,morenob},  and the  formation  time distribution  of
dark matter haloes \citep{st04,gmst07}.

The extended Press-Schechter  formalism can be numerically implemented
to  produce stochastic  realizations of  the merging  history  tree of
haloes  of any  mass; these  Monte Carlo  merging histories  can have,
theoretically, arbitrary resolution in mass and time, and their result
is the  full population  of progenitor haloes  at all times,  for some
final halo. The only thing  they do not provide is spatial information
on  the  haloes themselves.   Therefore,  when  the  focus is  on  the
statistical properties of haloes and not on their position or internal
structure, Monte Carlo merger trees have the advantage - over $N$-body
simulation -  of virtually  unlimited mass resolution.  Implementing a
code to reach the  required resolution is however not straightforward,
and caution  must be used in  order to preserve  consistently with the
theory.

In this  paper we will build  a spherical collapse  Monte Carlo merger
tree with arbitrary mass and time-step resolution, and will consider a
$\mathrm{\Lambda  CDM}$-power spectrum. We  will populate  a Milky-Way
sized halo with subhaloes with masses as small as $10^{-6} M_{\odot}$.
This value correspond  to the typical Jeans mass for  the a CDM Weakly
Interacting Massive  Particle (WIMP) particle  with $m_{\rm DM}  = 100
GeV$  \citep{ghs04,ghs05}.  Such  a  value for  the  minimum mass  can
actually vary between $10^{-12}$  and $10^{-4} M_{\odot}$ depending on
the underlying particle physics \citep{profumo}.

We will take  into account the progenitor haloes accreted by the main
branch of  our Milky-Way like halo,  in order to obtain  a snapshot of
the  spatial  distribution  of  substructures today.   
{\bf  In this way we will select only the first order of substructures which populate
the halo today (those which are orbiting in the host halo potential), 
in order to correctly model the radial dependence of subhalo properties.
In other words, we will avoid the bias of our previous works (see,e.g. \cite{gpt08})
towards the small scale structures. In those works, the embedding of sub-subhalos within subhalos 
was not treated correctly. Each structure existing inside the main halo
was indeed considered as orbiting inside the potential of the host halo itself. 
In this way, an analytical treatment
of the effect of subhalos on the expected photon flux from DM annihilation was possible. However,
it was not possible to take into consideration the fact that a number of small scale structures
are indeed sub-substructures, that is to say they are orbiting inside the potential of the subhalo they
belong to, and not inside the potential of the host halo. 
In the present work we will be able to separate and study the effect of introducing 
sub-subhalos.} 
With  the  same partition  code we will thus compute the  subhalo population  of subhaloes
(i.e. subhaloes  within subhaloes), and  we will consider  the special
case of  a Draco-like satellite.  In  the second part of  the paper we
will estimate the $\gamma$-ray emission from dark matter annihilation,
in  the  subhalo population  of  the  Milky-Way  and of  a  Draco-like
subhalo, and will discuss  the different contributions to $\gamma$-ray
emissions due  to the smooth  and clumpy components of  the considered
systems.

The paper is organized as  follows: in Sec.~\ref{tree} we describe the
merger tree technique developed by \citep{sl99} and its generalization
to a  $\mathrm{\Lambda CDM}$ power  spectrum.  Sec.~\ref{partition} is
dedicated to the  study of the merger tree of  a Milky-Way sized halo,
and to  compute its mass accretion  history along cosmic  time and its
satellite   mass  function.  In   Sec.~\ref{secdraco}  we   model  the
hierarchical growth  of a sample of Draco-like  satellites until their
merging   time  with   the   main  Milky-Way   progenitor  halo.    In
Sec.~\ref{subgamma}  we  study  the  $\gamma$-ray emission  from  dark
matter annihilation  in subhaloes  and sub-subhaloes and  estimate the
possibility  to   be  detected  with  the   Fermi-LAT  telescope.   In
Sec.\ref{deconc} we discuss our results and conclude.

\section{Method and Merger Tree}
\label{tree}

The simplest algorithm for a merger  tree uses a binary split. In this
scenario each halo is split in two haloes at an earlier epoch. Each of
these is in turn divided in other two pieces, and so on until all halo
masses  fall below  an arbitrary  chosen and  desired  mass resolution
$m_{\epsilon}$     \citep{lc93,colekaiser88,cole91,kw93}.     However,
\citet{sk99} showed  the failure  of using a  binary merger  tree.  In
this way the  first halo which is chosen from  the Press \& Schetchter
distribution follows  the correct probability,  but the second  one is
chosen  only  in  order to  conserve  mass  and  does not  follow  the
theoretical  model as  expected.   This leads  to  a conditional  mass
function and formation redshift  distribution in disagreement with the
extended  Press  \& Schechter  predictions.   To  solve this  problem,
\citet{sk99}  proposed   a  new   algorithm  able  to   reproduce  the
conditional  mass function  quite well,  at  the expenses  of using  a
finely tuned grid of time-steps.

On the  other hand, \citet{sl99}  used the results  of \citet{sheth96}
and  realized  that,  for  white-noise  initial  conditions,  mutually
disconnected  regions are mutually  independent.  In  this case  it is
possible to  split a halo  into progenitors very efficiently.   The so
obtained conditional mass function  is in excellent agreement with the
theoretical model  of the spherical collapse.  The  great advantage of
this  method is  that it  is possible  to obtain  arbitrary  high mass
resolution for  any given time-step, generating progenitors  in a very
fast and efficient way.

In  this section  we will  describe the  way the  method  developed by
\citet{sl99} can  be generalized considering  a $\mathrm{\Lambda CDM}$
power spectrum Using  this algorithm, we will follow  the creation and
the assembly history of a present-day Milky-Way sized halo.

Assuming that the dark matter is a WIMP, the Jeans mass of the smaller
DM halo is given  by its free-streaming scale \citep{ghs04,ghs05}, and
for  a DM  particle mass  of 100  GeV this  smallest halo  mass  has a
typical  mass $m_{J}  =10^{-6}\,M_{\odot}$.  We  therefore  extend our
merger tree down to microsolar  mass resolution, in order to study the
present-day subhalo population down to the dark matter Jeans mass.

\subsection{Poissonian Distribution and Gaussian Initial Conditions}

Let  us  start  considering  an  initial  Poissonian  distribution  of
identical  particles. \citet{epst83}  and \citet{sheth95}  showed that
the probability to find a  clump containing $N$ particles is expressed
by the Borel distribution \citep{borel42}:
\begin{equation}
\eta(N,b) = \frac{(N b)^{N-1} \mathrm{e}^{-N b}}{N !}\,,
\label{boreleq}
\end{equation}
where the variables $N$  and $b$ are such that $N \ge  1$ and $0 \le b
<1$.  \citet{sheth95}  also showed that this equation  can be extended
to  the  continuous  case,  in  order to  describe  dark  matter  halo
clustering. In  this case the  variable $b$ has a  redshift dependence
given by:
\[
b = 1/(1 + \delta_{c}(z))\,.
\]
where $\delta_c(z)$ is the critical overdensity threshold predicted by
the spherical collapse model, decreasing as the universe expands.  The
probability that a randomly chosen particle belongs to an $N$-clump is
given by  $f(N,b) = (1 - b)N  \eta(N,b)$. For large values  of $N$ and
small values  of $\delta_{c}$, the  factorial Stirling's approximation
gives:
\begin{eqnarray}
f(N,\delta_c) & = &  \frac{\delta_{c}}{(1 + \delta_{c})} \left(
  \frac{N}{1       +      \delta_{c}}       \right)^{N       -      1}
  \frac{\mathrm{e}^{-N/(1+\delta_{c})}}{(N  -  1)!}   \nonumber  \\  &
  \rightarrow  &  \frac{\delta_{c}}{\sqrt{2  \pi  N}}  \exp  \left(  -
  \frac{N \delta^2_{c}}{2}\right)\,. \label{s95e}
\end{eqnarray}
Since   all   matter   is   in   clumps,   it   holds   the   relation
$\sum_{N=1}^{\infty}f(N,\delta_c)=1$. \citet{sheth95} showed also that
Eq.  (\ref{s95e}) can be  obtained starting  from an  initial Gaussian
density field  with white noise initial conditions  ($P(k) \sim k^{n}$
with  $n=0$).    This  equivalence  underlines   that  the  Poissonian
distribution  can be  thought as  the analogue  discrete of  the white
noise Gaussian power spectrum \citep{bondetal91,lc93,lc94}.

The critical collapse overdensity  decreases with the expansion of the
universe,  so that  small systems  collapse earlier  than  large ones,
i.e.,  dark  matter  halo  clustering progresses  hierarchically.   An
important  quantity  that describes  the  hierarchical  growth of  the
haloes is the conditional  distribution. It gives the probability that
a particle, belonging to a clump  with $N$ particles at time $b_0$, is
part of an $n$-clump at $b_1<b_0$, and can be written as:
\begin{eqnarray}
  f(n,b_1|N,b_0) & = &  N \left(1 - \frac{b_1}{b_0} \right) \binom{N}{n}
  \frac{n^n}{N^N}   \nonumber  \\   &  \times   &  \left(\frac{b_1}{b_0}
  \right)^{n-1} \left[N - n\frac{b_1}{b_0} \right]^{N-n-1}\,,
\end{eqnarray}
where $1 \le n \le N$ and $0 \le b_1/b_0 \le 1$ \citep{sheth95}.

However, for a complete description  of the merging history tree of an
$N$-clump  at $b_0$, we  also need  to know  the probability  that, at
$b_1$,  it  is  divided in  a  sample  of  $n_j$ $j$-clumps  with  $k$
subfamilies (so that  $n_1+...+n_N=k$).  Recalling the conservation of
the  particle  number,  $\sum_{j=1}^k  j  n_j  =  N$,  the  Poissonian
Galton-Watson branching process gives  the following equation for this
probability:
\begin{eqnarray}
  p(n_1,...,n_k,b_1|N,b_0)    &    =    &   \frac{[N    (b_1-b_0)]^{n-1}
    \mathrm{e}^{-N(b_1-b_0)}}{\eta(N,b_0)}  \nonumber   \\  &  \times  &
  \prod_{j=1}^{k} \frac{\eta(j,b_1)^{n_j}}{n_j!}\,,
\label{gwbpeq}
\end{eqnarray}
where $\eta(l,b)$  is the Borel  distribution with time  parameter $b$
(see \citet{sheth96} for more details).  Since $b_0$ can be related to
a density, the volume of the $N$-clump can be written as:
\begin{equation}
V_{N,0} = \frac{N}{\bar{n}(1+\delta_{c,0})}\,,
\end{equation}
where  $\bar{n}$   denotes  the  average  density   of  the  universe.
Eq.~(\ref{gwbpeq}) can be thought of  as the probability that a region
with $N$ particles, having density $\bar{n}(1+\delta_{c,0})$, contains
$k$  subregions  each  at average  density  $\bar{n}(1+\delta_{c,1})$,
where $\delta_{c,1} \ge \delta_{c,0}$.

The  merging history  of a  present-day  halo is  described using  the
Extended-Press \&  Schechter formalism \citep{lc93}  and formulated in
terms  of  its  conditional   mass  function  along  consecutive  time
steps. Let us  now consider an $n$-subclump of  the $N$-clump, at time
$b_1$; in Appendix  A of \citet{sl99} it is shown  that, if $V_{n,1} =
n/\bar{n}(1+\delta_{c,1})$  is its  associated  volume, the  remaining
particles $N-n$ will occupy a volume such that:
\begin{equation}
\frac{N-n}{V_{N,0}  -   V_{n,1}}  \equiv  \bar{n}   (1+\delta'_{c})  =
\frac{\bar{n}}{b'}\,,
\end{equation}
where $b'1/(1+\delta'_{c})$  is the unknown quantity  to be determined
and represents the density in the volume $V_{N-n}$.

The spherical collapse model  predicts that a group of uncollisionless
$N$ particles,  with the same  mass $m_{\rm DM}$, collapses  forming a
dark  matter $M$-halo  ($M= N  \times  m_{\rm DM}$),  if the  smoothed
density  fluctuation  filed on  scale  $R  \sim  M^{1/3}$ exceeds  its
predicted  critical  virial  value  \citep{eke}.  For  each  collapsed
system we can define its associated mass variance as:
\begin{equation}
  S(M) = \frac{1}{(2 \pi)^3 V_{R}} \int \mathrm{d}^3 k
  \langle | \hat{\delta}(k) |^2 \rangle \hat{W}^2(kR)\,, \label{massv}
\end{equation}
where $\hat{W}(kR)$  and $\hat{\delta}(k)$ represent  respectively the
Fourier transform of the smoothing  window function and of the density
fluctuation field.

It   is   possible   to    show   that   in   the   continuous   limit
Eq.s~(\ref{boreleq}) and (\ref{gwbpeq}) give:
\begin{equation}
  f(\nu)  \mathrm{d} \nu  = 2 \sqrt{\frac{1}{2 \pi}}  \mathrm{e}^{- \nu^2/2}
  \mathrm{d}\nu\,. \label{gauss}
\end{equation}
Taking  $\nu=\delta_c/\sqrt{S}$, this equation  describes, at  a fixed
redshift (and  so at  a given $\delta_{c}$),  the number  of collapsed
systems  with mass  variance  between $S$  and  $S+ \mathrm{d}S$.  The
factor  $2$  takes  into  account the  so-called  \emph{cloud-in-cloud
  problem},  which is  the possibility  that at  a given  instant some
object, which is nonlinear on scale $M$, can be later contained within
another object,  on a larger mass scale.   Eq. (\ref{gauss}) describes
also the conditional mass function at time $\delta_{c,1}$, considering
a  halo  at  time   $\delta_{c,0}$  with  mass  variance  $S_0$,  when
$\nu=(\delta_{c,1}-\delta_{c,0})/\sqrt{s-S_0}$.
\begin{figure*}
  \centering
  \includegraphics[width=\hsize]{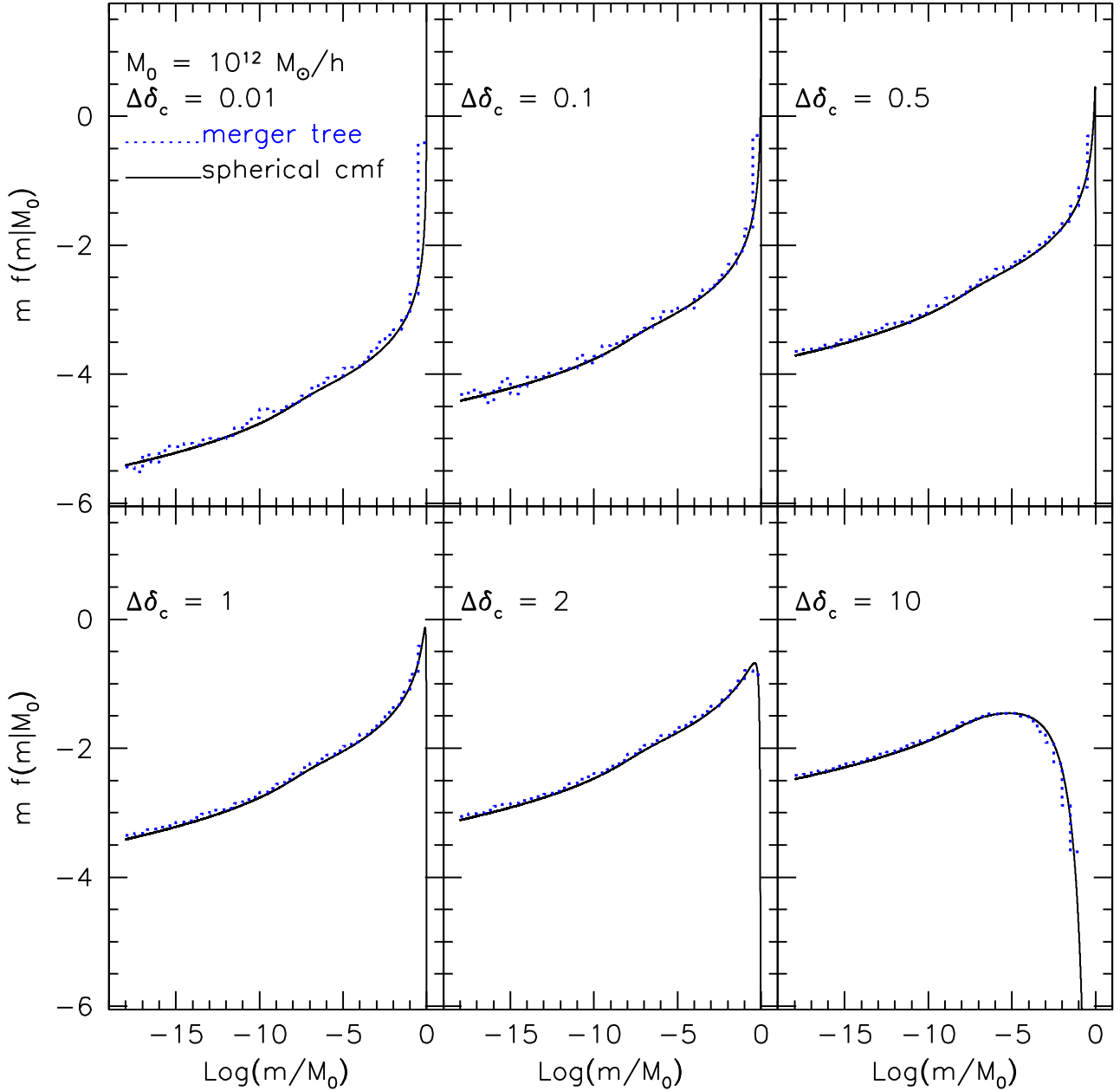}
  \caption{Spherical  collapse   conditional  mass  function   at  six
    different redshifts, obtained using  the SL99 tree.  We considered
    a  $\mathrm{\Lambda  CDM}$  power  spectrum extrapolated  down  to
    $10^{-6}\,M_{\odot}$  as  in  \citet{gpt08}.   In each  panel  the
    histogram shows the average of $10^4$ realizations of partitioning
    an  $M_0$-halo. The solid  line represents  the prediction  of the
    spherical  collapse   model  at  the   considered  redshift.   The
    partition was performed  both with a single and  a multi-step: the
    distributions are equivalent.\label{cmffig01}}
\end{figure*}

Before taking  into account  a $\mathrm{\Lambda CDM}$  power spectrum,
let us consider the simpler  case of a white-noise power spectrum.  We
recall that  when $P(k)  \sim k^n$, Eq  (\ref{massv}) becomes  $M \sim
S^{-(n+3)/3}$, that for  $n=0$ gives $M \sim 1/S$.   Let us suppose we
want  to split an  $M_0$-halo, at  time $\delta_{c,0}$,  in progenitor
haloes at $\delta_{c,1}$.  To generate  the first progenitor we draw a
random  number  $\tilde{\nu}_1$  from  the Gaussian  distribution,  Eq
(\ref{gauss}), and compute its associated mass variance, $\tilde{s}_1$
considering that
\begin{equation}
  \tilde{\nu}_1             =            \frac{\delta_{c,1}            -
    \delta_{c,m_{\epsilon}}}{\sqrt{\tilde{s}_1-s_{m_{\epsilon}}}}\,,
\label{sleq1}
\end{equation}
where for this  first progenitor $m_{\epsilon}=M_0$ ($s_{m_{\epsilon}}
\sim  1/m_{\epsilon}$) and  $\delta_{c,m_{\epsilon}}  = \delta_{c,0}$.
Its  physical mass,  $\tilde{m}_1$, can  be directly  computed  as the
inverse of its mass variance.   Since for a white-noise power spectrum
disconnected volumes are mutually  independent, the overdensity of the
remaining mass $m_{\epsilon} = M_0 - \tilde{m}_1$ will be given by the
volume conservation relation in the continuous limit:
\begin{equation}
\delta_{c,m_{\epsilon}}=\delta_{c,1}-
\frac{(\delta_{c,1}-\delta_{c,0})}{m_{\epsilon}/M_0}\,.
\label{sleq2}
\end{equation}
To generate the  second halo we draw another  number from the Gaussian
distribution,  compute the mass  variance from  the Eq.~(\ref{sleq1}),
and hence derive the corresponding  mass.  The remaining mass will now
be   $m_{\epsilon}=M_0  -   \tilde{m}_1  -   \tilde{m}_2$,   with  the
corresponding  overdensity given  always  by Eq.~(\ref{sleq2}).   Keep
going  with this  procedure it  is possible  to generate  a  sample of
progenitor haloes until the desired $m_{\epsilon}$ mass resolution.

\subsection{$\mathrm{\mathbf{\Lambda}}$CDM Power Spectrum}

In  the case  of  a general  power  spectrum the  algorithm should  be
modified.   The  assumption  that  disconnected volumes  are  mutually
independent does  not hold anymore when the  initial conditions differ
from  white-noise.   Despite  this,  \citet{sl99} noticed  that,  when
expressed as a  function of the variance rather than  of the mass, all
excursion  set quantities are  independent of  the power  spectrum. In
this framework, each chosen mass  $\tilde{m}$, can be treated not as a
progenitor  having a  mass  $\tilde{m}$,  but as  a  region of  volume
$\tilde{v}$  containing  a  mass  $\tilde{m}$,  populated  by  $\zeta$
objects having all the same mass $\mu$, with $\zeta = \tilde{m}/\mu$.

The number of objects is obtained by requiring that they have the same
mass variance,  that is $s(\mu)=1/\tilde{m}$.  For  a scale-free power
spectrum  $P \sim  k^n$ ($\alpha  = n  + 3/3$)  $\zeta =  m^{(\alpha -
  1)/\alpha}$, and for $n=0$ we have $\zeta=1$, the region $\tilde{v}$
contains exactly  one halo, as seen  in the previous  section.  For $n
\ne 0$ and  general power spectrum, $\zeta$ is  neither unity nor even
integer.  However, we  will show in the next  section that considering
$\zeta =  \mathrm{NINT}(\tilde{m}/\mu)$ (i.e. considering  the nearest
integer  to the  mass  ratio)  the progenitor  mass  functions are  in
excellent agreement  with the theoretical prediction  at all redshifts
and down to $m_{\epsilon}=m_J=10^{-6}\,M_{\odot}$.

We recall  that the $\mathrm{\Lambda  CDM}$ and the  white-noise power
spectrum should satisfy the relation:
\begin{equation}
s_{\rm wn}(M_0) = s_{\mathrm{\Lambda CDM}}(M_0)\,,
\label{wnlcdmpower}
\end{equation}
where $M_0$ is the initial mass  to be split. This guarantees that for
the considered initial mass there is  one halo in the volume $V_0$ for
both power spectra.

\section{Partition of a Milky-Way size halo with micro-solar mass resolution}
\label{partition}

Let us now take into account the case of a $\mathrm{\Lambda}$CDM power
spectrum.     The     density    parameter    and     mass    variance
($\Omega_{\Lambda}=0.7,\,\Omega_{m}=0.3$  and  $\sigma_8=0.772$)  have
been  chosen  to  agree  with  the recent  3-year  WMAP  data  release
\citep{spergel07}.   We have linearly  extrapolated the  mass variance
down to the  $m_{J}=10^{-6}\,M_{\odot}$ integrating the power spectrum
using a top-hat  filter in the real space  \citep{gpt08}.  In order to
have  one  physical  mass  both  for  the  $\mathrm{\Lambda  CDM}$  and
white-noise  power spectrum,  we should  consider a  white-noise power
spectrum normalized such that Eq.~(\ref{wnlcdmpower}) holds.  For this
reason, because we think in term of the mass variance than the fisical
mass, the mass  resolution $m_{\epsilon}$, for a $\mathrm{\Lambda}$CDM
power spectrum,  corresponds to the mass  resolution $m_{\epsilon, \rm
  wn}$      for      the      white-noise     one,      such      that
$s_{\mathrm{wn}}(m_{\epsilon,\rm      wn})=s(m_{\epsilon})$,      with
$m_{\epsilon,\rm wn}>m_{\epsilon}$.

\subsection{The main branch and the satellite mass function}

Let  us   consider  a  present-day   Milky-Way  sized  halo   ($M_0  =
10^{12}\,M_{\odot}/h$)  and suppose we  wish to  generate a  sample of
progenitors at different redshifts down to $m_{\epsilon} \equiv m_{J}=
10^{-6}\,M_{\odot}$,  using a  $\mathrm{\Lambda CDM}$  power spectrum.
Such mass  resolution corresponds to  $m_{\epsilon,\rm wn}=2.15 \times
10^{10}\,M_{\odot}/h$.   We  proceed  as  described  in  the  previous
section  and   obtain  the   conditional  mass  function   plotted  in
Fig.~\ref{cmffig01}.    The  number  of   progenitor  haloes   in  the
$\mathrm{\Lambda CDM}$ power  spectrum, for a given value  of the mass
variance, was obtained  by computing the nearest integer  of the ratio
$\tilde{m}/\mu$,       where       $s_{\mathrm{wn}}(\tilde{m})       =
s_{\mathrm{\Lambda  CDM}}(\mu)$.  In each  panel the  dotted histogram
shows the result of averaging  $10^4$ realizations, and the solid line
shows the  theoretical prediction  from the spherical  collapse model.
At each  redshift we  generated progenitors both  with a single  and a
multi-steps technique  starting from $z=0$.  Since the SL99  method is
independent on the time-step, the  results from the two different ways
of progress are in perfect agreement.

In  order to generate  the merging  history tree  along the  halo main
branch, we ran a sample of $10^4$ white-noise tree realizations for an
$M_0=10^{12}\,M_{\odot}/h$  initial  halo  and  a mass  resolution  of
$0.0215  \times  M_0$.   As  stressed  by SL99  (and  demonstrated  in
Fig.\ref{cmffig01}), this method is time-step independent: to obtain a
fine enough description of the  merging of satellite haloes we chose a
time resolution $\Delta \delta_c = 0.01$.  We followed the main branch
(the  most  massive  halo  among the  progenitors)  along  consecutive
time-steps, storing all the  information about the accreted progenitor
haloes (termed  ``satellites'', as in \citet{gtv}), until  the mass of
the  main branch becomes  as small  as the  mass resolution.   We call
merging redshift $z_m$ the most  recent redshift before a satellite is
incorporated in the main halo.

Finally,  the poissonian  trees is  converted in  the $\mathrm{\Lambda
  CDM}$ one as explained in the previous section.  We recall that each
progenitor halo in the first  tree is converted into $\zeta$ haloes in
the  second, in  order to  conserve the  mass variance  in  both power
spectra.

\subsection{Unevolved subhalo population}

\begin{figure}
  \centering
  \includegraphics[width=\hsize]{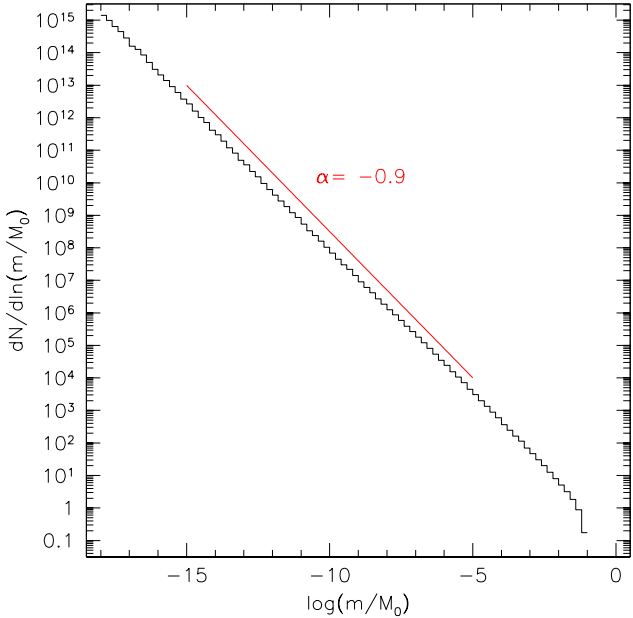}
  \caption{Satellite  mass function for  a present-day  Milky-Way size
    halo. The histogram shows the result of $10^4$ realizations of the
    merger history  tree. The  mass of each  satellite is  rescaled in
    units of the final host halo mass.  The solid line shows the slope
    of the least squares fit to the histogram down to $10^{-5}M_0$.
   \label{maccrmontc}}
\end{figure}

\citet{gpt08}  studied the  substructure population  of  a present-day
Milky-Way  halo  considering  its  progenitor mass  functions  at  any
redshift $z>0$,  and assuming that all progenitor  haloes survived and
retained  their  original  virial   mass  until  redshift  zero.  This
analytical   distribution  for  the   progenitor  mass   function  was
extrapolated down to microsolar mass  resolution in order to study the
$\gamma$-ray  emission  from   all  substructures  that  populate  the
galactic  halo.   Integrating the  conditional  mass  function at  any
redshift in  different mass bins, the full  hierarchy of substructures
was  automatically  taken  into  account: subhaloes  within  subhaloes
within subhaloes, and so on.  The $\gamma$-ray emission process due to
dark  matter annihilation  is proportional  to  the square  of the  DM
density, and  the possibility that subhaloes might  boost the expected
flux   was  taken  into   consideration  (see   \citet{gpt08,pbb}  and
references therein).

In  the   present  work  we  compute  the   $\gamma$-ray  emission  by
considering as substructure only the progenitor haloes accreted by the
main branch of  our merger tree. \textbf{This procedure  does not take
  directly into account the whole substructure population but only the
  first hierarchy of the population.  Progressing in} this way, we can
isolate the  contribution to the  boost from the single  DM structure,
including  its own  sub-substructures  \textbf{subsequently, following
  its  merging history  tree}.  Moreover,  in  this way  we can  model
correctly the spatial satellite  distribution inside the Galaxy, since
many small mass halos will be in fact embedded inside larger ones.

In  Fig.~\ref{maccrmontc} we  plot the  satellite mass  function, i.e.
the mass function  of progenitor haloes accreted directly  by the main
halo progenitor  at any redshift.   Satellite masses are  expressed in
units  of the  present-day mass  of the  host  halo \citep{vtg05,gtv}.
This figure  shows that the mass  distribution is well  described by a
single  power law $\mathrm{d}N/\mathrm{d}\ln(m)  \propto m^{-\alpha}$,
with  slope  $\alpha=-0.9$,  plus  an  exponential cut  off  at  large
masses. The slope was obtained  computing the least squares fit to the
data down  to $m = 10^{-5}M_0$.   This slope is  slightly steeper than
that obtained  by \citet{gtv}  using numerical $N$-Body  simulations (
$\alpha_{\mathrm{sim}}\approx -0.8$).  Such  a discrepancy is probably
due to the fact that the present merger tree is based on the spherical
collapse model, while numerical simulations are better described using
ellipsoidal collapse \citet{smt01,st02,gmst07}.

The  satellite  mass function  in  Fig.~\ref{maccrmontc} was  obtained
using  the   Monte  Carlo  code  described   in  Sec.~\ref{tree}.   We
considered a  halo of  mass $M_0$ at  redshift $z=0$ and  followed the
main branch  of its merging  history tree back  in time; we  then: (i)
generated     a     sample      of     progenitors     at     redshift
$\delta_{c,1}=\delta_{c,0}  + \Delta  \delta_c$;  (ii) identified  the
main (most  massive) progenitor halo; (iii) re-run  the partition code
computing its  progenitors at redshift $\delta_{c,2}  = \delta_{c,1} +
\Delta \delta_c$.  We iterated these  steps until the mass of the main
progenitor halo dropped below the mass resolution $m_{\epsilon}$. This
produced  the  first  merging  history  tree. We  repeated  the  whole
procedure,  starting from the  same mass  $M_{0}$, to  generate $10^4$
histories, to be averaged on.

After   computing   the   satellite   mass   function   (also   called
\emph{unevolved} subhalo  mass function) we associated to  each halo a
concentration      parameter     related      to      the     quantity
$\nu(z_m,m)=\delta_{c}(z_m)/\sqrt{s(m)}$, where $z_m$ is the satellite
merging redshift onto the host halo, and $m$ its virial mass at $z_m$.

\begin{figure}
  \centering
  \includegraphics[width=\hsize]{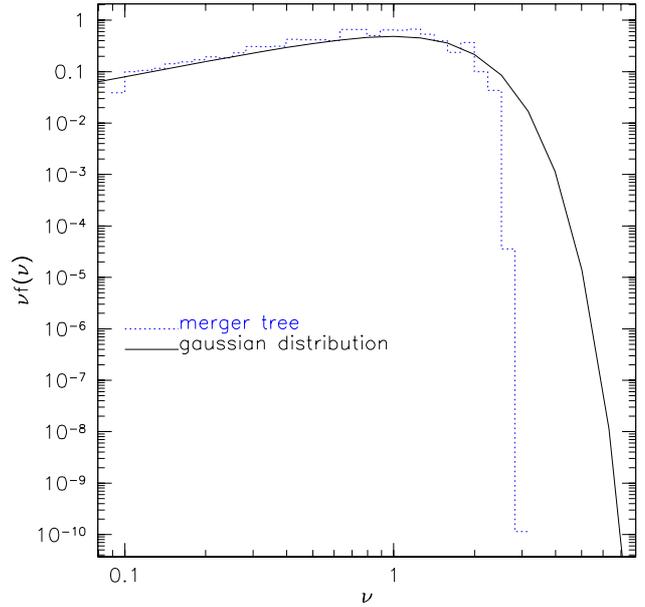}
  \caption{Satellite mass  function in term of  the rescaled parameter
    $\nu(z_m,m)=\delta_c(z_m)/\sqrt{s(m)}$.  The dotted histogram show
    $10^4$ realizations of  the merger tree, while the  solid curve is
    the Eq.~(\ref{gauss}).\label{satnufun}}
\end{figure}

By definition, $\nu$ defines the rareness of the density peak the halo
belonged  to  at  the  epoch  of  merging.  Higher  peaks  of  density
fluctuations   correspond   to   a   collapse   happened   at   higher
redshifts. Such a  rare halo will therefore be  more concentrated than
the bulk of halos with the same mass which formed at later epochs.

\citet{dmm05} used  $N$-Body simulations to show  that the present-day
subhalo distribution preserves memory of their initial conditions.  In
detail, high density peaks are found to be more centrally concentrated
and  to  move   on  more  eccentric  orbits  than   the  overall  mass
distribution.  This correlation  has been interpreted and parametrized
by  \citet{dmm05}  using  the  variable  $\nu=\delta_{c}(z)/\sqrt{s}$.
This variable (i)  is related to the subhalo  concentration - which in
turn  determines the $\gamma$-ray  emission, and  (ii) enables  one to
compute the  spatial distribution  in the host  halo (see Eq.   (1) of
\citet{dmm05}).   The   details   of   the   models   are   given   in
Sec. \ref{rgammamerging}

In Fig.\ref{satnufun} we plot the  satellite mass function in terms of
the universal  variable $\nu$.  The dotted histogram  shows the result
of $10^4$ Monte  Carlo realizations of the partition  algorithm (as in
Fig.\ref{maccrmontc}),   while  the  solid   curve  is   the  Gaussian
distribution, Eq.~(\ref{gauss}).

The figure shows that the satellite mass function is well described by
a  gaussian distribution  for small  and intermediate  masses,  with a
cutoff  at $\nu$=2.   This  fact can  be  qualitatively understood  by
recalling   that  the   progenitor  mass   function  has   a  Gaussian
distribution when expressed in  the rescaled variable $\nu$.  For each
evolutionary  step  $\mathrm{d}z$,  the  satellite  mass  function  is
obtained  by  removing from  all  progenitors  the  most massive  one.
Integrating  over  redshift,  the  total satellite  mass  function  is
effectively  an integral  over different  gaussians deprived  of their
most massive progenitor.

\section{Draco}
\label{secdraco}

The algorithm described  in the previous sections can  also be used to
study and constrain  some properties of the dark  matter halos hosting
the Milk-Way satellite galaxies.  As an example let us consider Draco,
a dwarf spheroidal galaxy with  an old stellar population, dark matter
dominated: Draco's  mass to light ratio  is estimated to  be $m/L \sim
100  M_\odot/L_\odot$  \citep{pryor92,irw95}.  This  makes  it a  very
interesting  target  to  investigate  the existence  of  dark  matter;
specifically,  we   study  the   possibility  of  an   observation  of
$\gamma$-ray due to DM annihilation inside Draco using the Fermi Large
Area Telescope (Fermi-LAT, formerly  known as GLAST).  To this extent,
we model the distribution of dark matter subclumps inside Draco.

\subsection{Mass and Merging Time Estimates}

Hierarchical  models of  galaxy  formation predict  that galaxies  are
hosted  by dark  matter haloes.   Such haloes  grow along  cosmic time
through repeated merging events.   In this scenario satellite galaxies
reside in  subhaloes accreted by the main  halo progenitor.  Satellite
haloes   grow   themselves    hierarchically,   accreting   mass   and
sub-progenitors up  to the  time when they  merge onto the  host halo.
After they fall in the  gravitational potential well of the host, they
start to loose mass due  to gravitational heating, tidal stripping and
close  encounters with  other subhaloes,  so that  only a  fraction of
their  initial  virial  mass  is  still self-bound  at  redshift  zero
\citep{vtg05,dkm07,dkm08,gtv}.

In order to  build the history tree for a  subhalo hosting a satellite
galaxy, we need to know its  virial mass at merging time $z_m$ and use
it as  our starting point.  From  its merging redshift we  then can go
backwards in time and reconstruct  its subhalo population, in order to
model its present-day sub-subhalo mass function and distribution.

\begin{figure}
  \centering
  \includegraphics[width=\hsize]{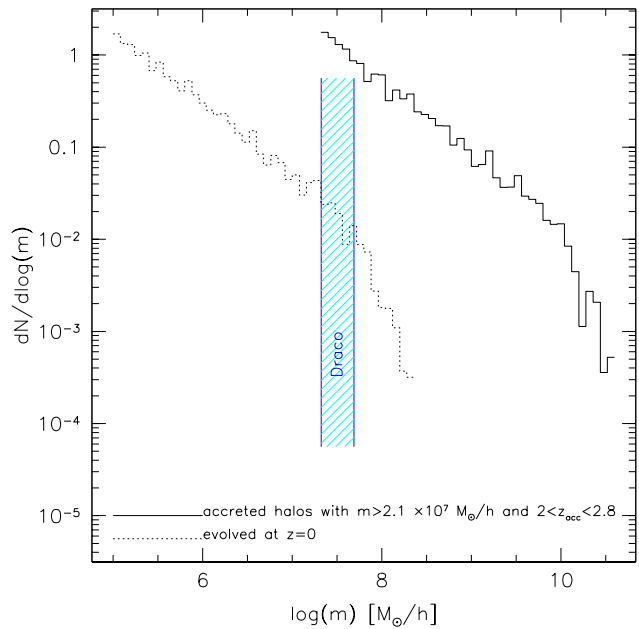}
  \caption{\label{dracop}\emph{Unevolved}  and  \emph{evolved} subhalo
    mass function  of Draco candidates  accreted by a  Milky-Way halo.
    The solid  histogram shows the satellite  distribution, the dotted
    one refers to  the present day mass function  evolved by using the
    average mass loss rate  derived by \citet{gtv}. The hatched region
    bounds the  most likely values  for the present-day mass  of Draco
    \citep{dracom1,dracom2}.}
\end{figure}

Using deep wide-field multicolor CCD photometry from the Sloan Digital
Sky Survey,  assuming a  \citet{king66} spherical model  of equivalent
size as  a reference and adopting a  line-of-sight velocity dispersion
of $10.7$ km/s \citet{aaddc} finds $3.5 \pm 0.7 \times 10^7 M_{\odot}$
within 28 arcmin while  \citet{dracom1}, considering $8.5$ km/2, finds
$2.2 \pm 0.5 \times 10^7\,M_{\odot}$ within 40 arcmin.  Differences in
the experimental mass  modeling determine the mass range  we allow for
the Draco-like  satellite in our analysis.  Considering  all the stars
within  the   tidal  radius,  \citet{dracom1}   determined  the  total
luminosity of the  Draco dwarf galaxy to be  $(L/L_{\odot})_i= 2.4 \pm
0.5 \times  10^5$.  \citet{dracom2}  also studied the  distribution of
dark  matter in  Draco by  modeling the  moments of  the line-of-sight
velocity distribution  of its stars from the  velocity dispersion data
of \citet{wilk06}, and obtained a  best-fitting total mass equal to $7
\times  10^7\,M_{\odot}$.  The  inferred mass-to-light  ratio  (in the
$V$-band) was $300\,M_{\odot}/L_{\odot}$, almost constant with radius.
On the other hand, we lack  a direct estimate for the merging redshift
of Draco onto the Milky  Way.  From its initial position and velocity,
the accretion time  of Draco can be derived  indirectly by considering
the  one-to-one  relation between  virial  radius  and accretion  time
implied   by  the  spherical   secondary  infall   model  \citep{sfm}.
\citet{draco} showed that an upper limit for Draco merging redshift is
$z_{m} \lesssim 2.8$.

We  have computed the  Draco virial  mass from  the previous  mass and
redshift estimates in  the following way. From each  one of the $10^4$
Monte Carlo merging trees of  a Milky-Way sized halo ($M_0$ = $10^{12}
\msun/h$) we noted down all satellites with mass larger than $3 \times
10^{7}\,M_{\odot}$  accreted  by the  main  branch  between $z=2$  and
$z=2.8$,  totaling  $8932$ Draco  candidates.   The redshift  interval
roughly corresponds to a time  interval of $1$Gyr from the upper limit
of the accretion  time computed by \citet{draco}.  The  lower limit on
the Draco  mass guarantees that  evolved masses will lie  today within
the  experimentally estimated  mass range.   In  Fig.~\ref{dracop} the
solid histogram  shows the mass  function of these  selected satellite
haloes.  Once inside  its host, each satellite will  loose mass due to
gravitational heating  and tidal stripping effects.   We modeled these
effects using  the results in Eq.  (10) of \citet{gtv}.  In that paper
the authors  measured the subhalo mass  loss rate in a  sample of high
resolution    $N$-body   haloes    ranging    from   $10^{11.5}$    to
$10^{15}\,M_{\odot}/h$.   Following their recipe,  the ratio  of $z=0$
self bound mass to the original virial mass of the subhalo is uniquely
determined  by the amount  of time  the subhalo  has spent  inside its
host.   The dotted  histogram in  Fig.~\ref{dracop} shows  the evolved
subhalo Draco-candidates mass function.  Considering that the measured
fractional mass loss rate by \citet{gtv} is independent of the subhalo
mass, the  evolved mass function has  the same slope  of the unevolved
one.

\subsection{The Sub-Tree}

\begin{figure}
  \centering
  \includegraphics[width=\hsize]{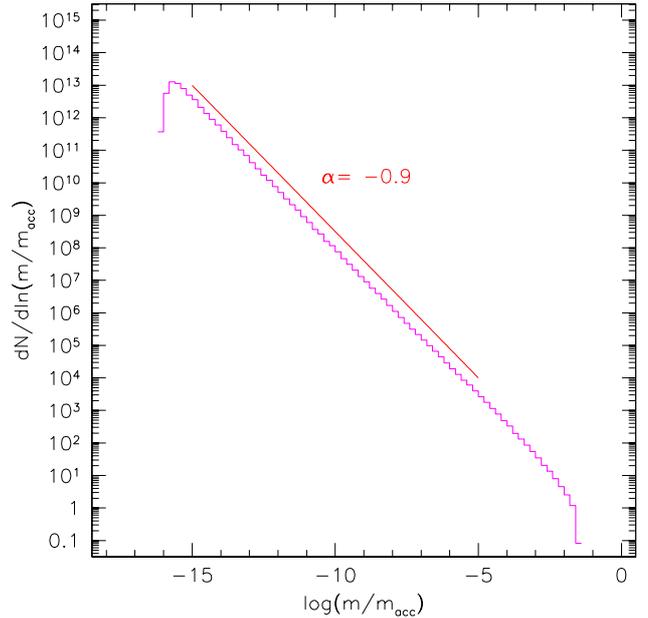}
  \caption{Satellite mass function accreted  by the main branch of the
    Draco-candidate  trees, until  the merging  on the  Milky-Way main
    progenitor  halo.  The  total number  of realizations  is $26796$.
    Comparing the histogram  with Fig.~\ref{maccrmontc} we notice that
    the  slopes  of  the  distributions  are identical.   This  is  in
    agreement  with  what  found  by  \citet{gtv,vtg05}  that  is  the
    \emph{unevolved}  subhalo  mass function  is  independent both  of
    final host halo mass and redshift.\label{dracoamf}}
\end{figure}

In order to compute the  Draco satellite mass function, each candidate
of our  sample has been  further evolved back  in time along  its main
branch  using the partition  code (as  it was  done for  the Milky-Way
halo) considering  the same  mass and time  step resolution.   In this
case each  merger history tree  starts at the  corresponding satellite
merging time.  To increase the statistical significance  of our result
we  run   three  realizations  for  each  of   the  $8932$  Draco-like
satellites, totaling $26796$ Monte Carlo merger tree realizations.  In
Fig.~\ref{dracoamf}  we  show  the \emph{unevolved}  sub-subhalo  mass
function accreted by  the Draco sample.  Comparing Fig.~\ref{dracoamf}
and Fig.~\ref{maccrmontc}, we see that the two accreted mass functions
are  indistinguishable: in  fact, as  found by  \citet{gtv,vtg05}, the
accreted mass  function is universal,  independent both of  final mass
and observation redshift. The slope  of the power law is again $\alpha
= -0.9$.

\section{$\gamma$-ray flux from Dark Matter annihilation in 
         sub(sub)structures}
\label{subgamma}

The photon  flux expected  from DM annihilation  in the  population of
galactic subhaloes can be modeled as
\begin{equation}
  \frac{d  \Phi_\gamma}{dE_\gamma}(E_\gamma,  \psi,  \Delta \Omega)  =  \frac{d
    \Phi^{\rm  PP}} {dE_\gamma}(E_\gamma) \times  \Phi^{\rm cosmo}(\psi,
  \Delta \Omega)
\label{flussodef}
\end{equation}
where $\psi$ defines the direction  of observation with respect to the
galaxy center, and $\Delta \Omega$ is the detector angular resolution.

The particle physics dependence in Eq.~\ref{flussodef} is given by the
annihilation spectrum and DM properties and is embedded in the term:
\begin{equation}
\frac{d   \Phi^{\rm   PP}}{dE_\gamma}(E_\gamma)   =  \frac{1}{4   \pi}
\frac{\sigma_{\rm   ann}  v}{2   m^2_\chi}   \cdot  \sum_{f}   \frac{d
  N^f_\gamma}{d E_\gamma} B_f\,,
\label{flussosusy}
\end{equation}
where  $m_\chi$  is the  DM  particle  mass,  $\sigma_{\rm ann}v$  the
self--annihilation cross--section  times the relative  velocity of the
two  annihilating  particles,  and  $d  N^f_\gamma  /  dE_\gamma$  the
differential  photon  spectrum  for  a  given  final  state  $f$  with
branching   ratio  $B_f$  which   we  model   after  the   results  of
\citep{fps04}.

In this paper we will set  $\Phi^{\rm PP} = 1$ and refer to $\Phi^{\rm
  cosmo}$ as  a scaled photon  flux.  In the following  subsections we
review different ways to compute $\Phi^{\rm cosmo}$ in the presence of
substructures.   We will  then show  the computation  relative  to the
merger tree technique presented in the previous sections and will show
the comparison between the different predictions.
 
\subsection{Results on the $\gamma$-ray flux from the analytical 
            description of the subhalo population}

\citet{pbb} have considered the existence of a population of
substructures  inside  a  DM  halo, described  by  the  mass
function
\begin{equation}  
{\rm d} N/{\rm d ln}(m) \propto m^{-1},
\label{mass}  
\end{equation}  
and  assumed that Eq.~\ref{mass}  describes the  mass function  of all
substructures in our  Galaxy as well as that  of sub-substructures, at
z=0. In this  case all the halos that ever accreted  onto the main one
and  survived   are  considered,  including  the   higher  level  ones
(substructures within  substructures).  Using this  prescription it is
then  impossible  to  discriminate  a sub-halo  from  a  sub-sub-halo,
affecting  in a wrong  way the  spatial distribution  of substructures
inside the main halo.

\citet{pbb} assumed that  the substructure spatial distribution traces
that of the underlying host mass  from $r_{vir}$ and down to a minimum
radius,  $r_{min}(m)$, within  which  disruption by  Galaxy tides  and
stellar encounters becomes relevant.

Folding these indications together  they modeled the number density of
subhaloes per  unit mass  at a distance  $R$ from the  Galactic Center
(GC) as:
\begin{equation}  
\rho_{sh}(m,R) = A  m^{-2} \frac{\theta (R - r_{min}(m))}{(R/r_s^{PH})
  (1 + R/r_s^{PH})^{2}} \msun^{-1} \kpc^{-3},
\label{rho}  
\end{equation}  
where $r_s^{PH}$ represents  the scale radius of the  parent halo (PH)
referred  to a Navarro,  Frenk and  White (NFW)  \citep{NFWp} profile.
The effect of tidal disruption  is accounted for by the Heaviside step
function $\theta (r - r_{min}(m))$.

To  determine the  tidal  radius, $r_{min}(m)$,  they  used the  Roche
criterion and compute it as  the minimum distance at which the subhalo
self-gravity  at  $r_s$ equals  the  gravity  pull  of the  host  halo
computed at the orbital radius of the subhalo.

To normalize Eq.~\ref{rho}, they imposed that $10\%$ of the PH mass is
distributed in subhaloes with masses in the range $[10^{-5}-10^{-2}] \
M_{PH}$.

Following the  previous prescriptions, they found about  $53\%$ of the
MW mass  ($M_{PH}= 10^{12}  \msun/h$, $r_s =  21.7 \kpc$,  $c_{200}$ =
9.8) condensed within $\sim  1.5 \times 10^{16}$ subhaloes with masses
in the range [$h \, 10^{-6},10^{10}] \msun/h$.

\citet{Petal08}  repeated the  calculation for  the Draco  Galaxy, for
which  $\sim$ 40\%  of  the  halo mass  ($M_{Draco}=  7 \times  10^{7}
\msun$,  $r_s =  0.4  kpc$,  $c_{200}$ =  21.2)  condensed into  $\sim
10^{12}$ halos with masses between [$h \, 10^{-6},10^{6}] \msun/h$.

For  each  substructure,  they   used  a  NFW  density  profile  whose
concentration parameter $c_{200}$ (referred  to the radius enclosing a
density  equal to  200 times  the critical  density) is  given  by the
\citet{Bullock01}  prescriptions for  the subhaloes  at  redshift zero
(assuming the subhaloes are  slightly more concentrated than the field
halos). The \citet{Bullock01} model holds  down to masses of the order
of $10^5  M_\odot$. In  order to cover  the whole subhalo  mass range,
they performed a double mass  slope extrapolation down to the smallest
masses, obtaining a concentration parameter of about 70 for a $10^{-6}
\msun$  halo at redshift  zero. This  value is  in agreement  with the
extrapolation  at  redshift  zero  of  what  found  in  the  numerical
simulations  by \cite{Diemand:2005vz},  who  isolated $10^{-6}  \msun$
halos at z=26.

Such  a  model  for   the  concentration  parameter  has  been  called
$B_{z_0,ref}$ model.

The  contribution  of  unresolved  substructures to  the  annihilation
signal along a cone of sight is given by
$$ \Phi^{\rm cosmo}(\psi, \Delta  \Omega) = \int_M \mathrm{d} m \int_c
\mathrm{d}  c \int \int_{\Delta  \Omega} \mathrm{d}  \theta \mathrm{d}
\phi \int_{\rm l.o.s} \mathrm{d} \lambda
$$
$$
[ \rho_{sh}(m,R(\rsun, \lambda,\psi, \theta, \phi)) \times P(c) \times
$$
\begin{equation}
\times  \Phi^{\rm cosmo}_{halo}(m,c,r(\lambda, \lambda  ', \psi,\theta
', \phi ')) \times J(x,y,z|\lambda,\theta, \phi) ]
\label{smoothphicosmo}
\end{equation}
where $\Delta  \Omega$ is the  solid angle of observation  pointing in
the  direction  of observation  $\psi$  and  defined  by the  detector
angular  resolution   $\theta$;  $J(x,y,z|\lambda,\theta,  \phi)$  the
Jacobian determinant;  $R$ the galactocentric  distance, which, inside
the  cone,  can  be  written  as  a function  of  the  line  of  sight
($\lambda$), the  solid angle ($\theta$  and $\phi$) and  the pointing
angle $\psi$  through the relation  $R = \sqrt{\lambda^2 +  \rsun^2 -2
  \lambda \rsun C}$,  where $\rsun$ refers to the  distance of the Sun
from  the Galactic  Center  and $C=\cos(\theta)  \cos(\psi)-\cos(\phi)
\sin(\theta)  \sin(\psi)$; $r$  is  the radial  coordinate inside  the
single subhalo  located at distance $\lambda$ from  the observer along
the line  of sight defined by  $\psi$ and contributing  to the diffuse
emission. Finally, $P(c)$ is  the lognormal distribution of the values
for the concentration parameters.  The expression
$$ \Phi^{\rm cosmo}_{halo}(m,c,r) = \int \int_{\Delta \Omega} d \phi '
d \theta ' \int_{\rm l.o.s} d\lambda '
$$
\begin{equation}
\left  [ \frac{\rho_{\chi}^2  (m,c,r(\lambda, \lambda  ',\psi,\theta '
    \phi '))} {\lambda^{2}} J(x,y,z|\lambda ',\theta ' \phi ') \right]
\, ;
\label{singlehalophicosmo}
\end{equation}
describes the emission from each subhalo and $\rho_\chi(m,c,r)$ is the
Dark Matter density profile inside the halo.

Numerical   integration  of   Eq.    \ref{smoothphicosmo}  gives   the
contribution  to  $\Phi^{\rm  cosmo}$  from  unresolved  clumps  in  a
$10^{-5} \sr$ solid angle along the direction $\psi$.

We  show the  results obtained  in this  theoretical framework  with a
dashed  line  in   Fig.  \ref{mw}  \citep{pbb}  for  the   MW  and  in
Fig.\ref{figdraco} \citep{Petal08} for the Draco galaxy.

\subsection{Results on the $\gamma$-ray flux from the analytical method 
         including the effect of the merging epoch}
\label{rgammamerging}

A further  degree of  detail was given  by \cite{gpt08},  who computed
Eq.~\ref{smoothphicosmo}  including  the  dependence  of  the  subhalo
spatial  distribution  from the  initial  conditions  when the  haloes
accreted  into  the  present-day  Milky  Way halo.  Such  a  model  is
characterized  by  the  universal  variable $\nu  (m)$.  According  to
\cite{dmm05} the DM density profile of our Galaxy can be written as
\begin{equation}
\rho_\chi(r)  = \frac{\rho_s}  {  \left( \frac{r}{r_s}  \right)^\gamma
  \left [  1 +  \left( \frac{r}{r_s} \right)^\alpha  \right]^{(\beta -
    \gamma)/\alpha}}
\label{density}
\end{equation}
with  $(\gamma,  \beta, \alpha)  =  (1.2,  3,  1)$. Subsequently,  the
following parameterization  is used to  reflect the fact  that material
accreted in areas with  high density fluctuations is more concentrated
toward the centre of the galaxy, and has a steeper outer slope:
$$
r_s \longrightarrow r_\nu = f_\nu r_s
$$
$$
f_\nu = \rm{exp}(\nu/2)
$$
\begin{equation}
\beta \longrightarrow \beta_\nu = 3 + 0.26 \nu^{1.6}
\label{nu_param}
\end{equation}
Including a step function to take into account tidal disruption, as in
Eq.  \ref{rho},  the number  density of subhaloes  per unit mass  at a
distance $r$ from the GC, for a given $\nu (m)$, becomes:
\begin{equation}  
\rho_{sh}(m,r,\nu) = \frac{ A m^{-2} \theta (r - r_{min}(m))} { \left(
  \frac{r}{r_\nu (m)} \right)^\gamma \left [ 1 + \left( \frac{r}{r_\nu
      (m)} \right)^\alpha \right]^{(\beta_\nu - \gamma)/\alpha}},
\label{rho}  
\end{equation}  
in  units of $\msun^{-1}  \kpc^{-3}$. The  mass dependence  in $r_\nu$
reflects    the   mass    dependence   of    the    virial   parameter
$r_s=r_{vir}/c_{vir}$.

\cite{gpt08} normalized the number of  subhaloes such that 10\% of the
MW  mass  is  distributed  in  subhaloes  with  masses  in  the  range
$[10^{-5}-10^{-2}]  \ M_{MW}$,  ending  up with  $2.4 \times  10^{16}$
subhaloes with  masses between $10^{-6} \msun$  and $10^{10} \msun/h$,
accounting for  74 \% of  the MW mass  ($M_{MW}= 1.4 \times  10^{12} \
\msun$, $r_s = 26 \kpc$).

The contribution to $\Phi^{\rm cosmo}$ is given by:
$$  \Phi^{\rm  cosmo}(\psi,  \Delta  \Omega)  =  \int_M  \mathrm{d}  m
\int_\nu  \mathrm{d} \nu \int  \int_{\Delta \Omega}  \mathrm{d} \theta
\mathrm{d} \phi \int_{\rm  l.o.s} \mathrm{d} \lambda \int_c \mathrm{d}
c
$$
$$  [ \rho_{sh}(m,R(\rsun,  \lambda,\psi, \theta,  \phi),  \nu) \times
  P(\nu(m)) \times P(c(m)) \times
$$
\begin{equation}
\times \Phi^{\rm cosmo}_{halo}(m,r(\lambda,  \lambda ', \psi,\theta ',
\phi '), \nu, c) \times J(x,y,z|\lambda,\theta, \phi) ]
\label{smoothphicosmoGPT}
\end{equation}
which accounts for the influence of cosmology in the flux computation.
$P(\nu(m))$  is the  probability  distribution function  for the  peak
rarity  $\nu(m)$,   calculated  using  the   extended  Press-Schechter
formalism.   $P(c(m))$ is the  lognormal probability  distribution for
$c$ centered  on $c_{vir}(m)$. While $P(\nu(m))$ is  determined by the
merging history  of each subhalo,  $P(c(m))$ describes the  scatter in
concentration for  haloes of equal mass \citep{Bullock01}  and the two
probabilities may  be assumed to  be independent.  As in  the previous
analytical estimate  of the subhalo population, no  distinction can be
made with this method between sub-halos and sub-sub-halos.

The  result  of  this  calculation  for  the  MW  is  shown  with  the
long-dashed line in Fig.~\ref{mw}.

\subsection{Results on the $\gamma$-ray flux from merger tree technique}
\label{rgammatree}

In this  work we have  described a merger  tree approach to  infer the
subhalo population of  both the MW and Draco.  In  this case, we don't
need to use the $P(\nu)$ and $\mathrm{d}N/\mathrm{d}m$ factors, nor to
apply any normalization.  Indeed, the  output of the merger tree gives
us directly the number of objects with a given mass and a given $\nu$,
which we call now $N(\bar m, \bar \nu)$.

The  total number  of sub(-sub)structures  found with  this  method is
$\sim 2.6 \times  10^{15}$ for the MW and $2.7  \times 10^{13}$ in the
case  of Draco, at  the merging  epoch.  In  the case  of the  MW, and
differently from  the analytical methods described  above, this number
represents  only the  subhaloes, while  sub-subhaloes must  be treated
separately, as it has been done for the Draco-like subhalo.

In order to compute the  $\gamma$-ray flux we can re-write the subhalo
distribution function as
$$
\rho_{sh}= A g(r,\nu(m)) f(m)
$$
so that
$$  N_{tot}=   \sum_{m_i}  N(m_i)  =   \int_{gal}  \mathrm{d}V  \int_M
\mathrm{d}m \int_\nu P(\nu(m)) \rho_{sh} \,
$$ which can be rearranged into the following expression:
$$   \sum_{m_i}  N_{m_i}=   \int_M  \mathrm{d}m   A   f(m_i)  \int_\nu
P(\nu(m_i)) \int_{gal} \mathrm{d}V g(r,\nu(m_i)) \,.
$$
Let's define 
$$
G(m_i)=\int_{gal} \mathrm{d}V  g(r,\nu(m_i)) \,. 
$$
The expression of $\Phi_{cosmo}$ can be then written as:
$$  \Phi^{\rm  cosmo}(\psi,  \Delta  \Omega)  =  \int_M  \mathrm{d}  m
\int_\nu  \mathrm{d} \nu \int  \int_{\Delta \Omega}  \mathrm{d} \theta
\mathrm{d} \phi \int_{\rm l.o.s} \mathrm{d}\lambda \int_c \mathrm{d}c
$$
$$  [ A  g(r,\nu(m))  f(m) P(\nu(m))  P(c(m)) \Phi^{\rm  cosmo}_{halo}
  \frac{G(m)}{G(m)} J] \,.
$$ 
With  respect  to  Eq.~\ref{smoothphicosmoGPT}  we have  only  written
$\rho_{sh}$ in an  explicit way and multiplied by  $G(m) \over G(m)$ =
1.   In this  way we  can  recognize and  use our  merger tree  output
$\sum_{m_i}  N_{m_i}$   in  the   following  way  ($\int_{gal}   dV  =
\int_{\Delta \Omega} d \theta d \phi \int_{\rm l.o.s} d\lambda$):
$$  \Phi^{\rm cosmo}(\psi,  \Delta \Omega)  =  \sum_{m_i} \int_{\Delta
  \Omega}   \mathrm{d}  \theta   \mathrm{d}   \phi  \int_{\rm   l.o.s}
\mathrm{d}\lambda \int_c \mathrm{d}c
$$
\begin{equation}
[N(m_i) g(r,\nu(M_i)) P(c(M)) \frac{1}{G(m_i)} \Phi^{\rm cosmo}_{halo} J \,.]
\label{smoothphicosmoMT}
\end{equation}
 
\begin{figure}
  \centering
  \includegraphics[width=\hsize]{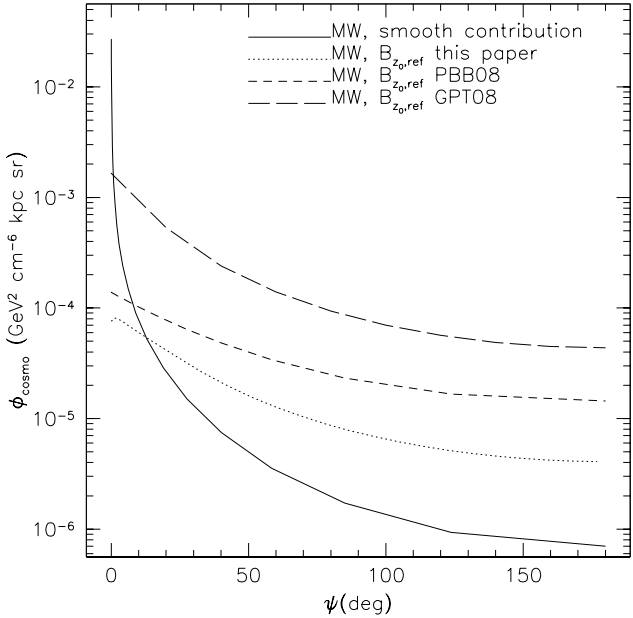}
\caption{ $\Phi_{cosmo}$  as a  function of the  angle of  view $\psi$
  from the galactic centre, computed for the MW.  }
\label{mw}
\end{figure}

The  result of  this calculation  is depicted  with a  dotted  line in
Fig.\ref{mw}.   We  note  that  the analytical  methods  predict  more
substructures,  and  therefore  a   higher  signal,  than  the  merger
tree. This  is mainly due to  the different mass  function slope (-2.0
versus -1.9).
 
In the case  of the Draco galaxy, the merger tree  was computed at the
Draco merging epoch (no sub-subhaloes accrete after merging), when its
mass was $9.78 \times  10^9 \msun$.  Although the sub-subhaloes evolve
in  redshift  and  loose  mass  as  the parent  halo  does,  for  each
sub-subhalo we considered a  NFW profile whose concentration parameter
has  been computed  for the  sub-subhalo  mass at  the merging  epoch,
motivated by the fact that  the inner profile of the structures should
not or poorly be affected by evolution.

From  the  observational point  of  view,  we  are interested  in  the
Draco-like galaxy as it is today, e.g.  with a mass content reduced by
a  factor  $\sim 10^{-2}$  due  to  tidal  interactions after  merging
\citep{gtv}.

To  estimate  the number  of  sub-subhaloes  in  Draco today  we  have
retained all  and only the subhaloes  which, at the  epoch of merging,
were inside the radius containing the Draco mass today, that is within
$r_c = 2.84$  kpc. $r_c$ is also similar to  the tidal radius obtained
using the  prescription given by \cite{Springel} to  obtain the number
of sub-subhaloes  today. The  total number of  subhaloes is  $6 \times
10^{-3}$ smaller than the initial one. This is the upper value for the
number of  subhaloes, since we are  not considering here  that half of
the subhaloes exit  the virial radius of the  parent halo during their
first  orbit \citep{bepi}.   In the  case of  field parent  halos they
would then be re-attracted inside the  halo, but if the parent halo is
a subhalo itself (like Draco) they would then be dispersed in the main
halo (that is the MW).

As  for $\rho_{sh}$,  we  have  used the  scale  quantities $r_s$  and
$\rho_s$ computed for the evolved  Draco-like galaxy, that is for a $7
\times 10^7 \msun$ halo.

\begin{figure}
  \centering
  \includegraphics[width=\hsize]{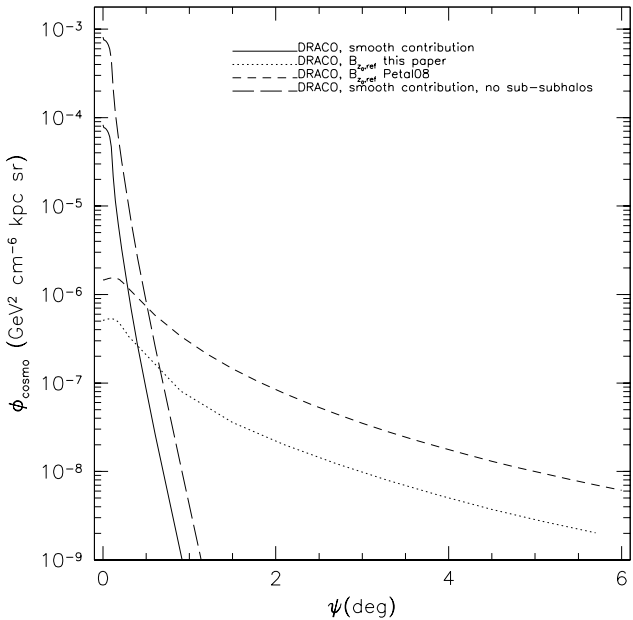}
  \caption{ $\Phi_{cosmo}$ as  a function of the angle  of view $\psi$
    from the galactic centre, computed for the Draco galaxy.}
\label{figdraco}
\end{figure}

The result  of the  computation of $\Phi_{cosmo}$  for Draco  is shown
with a dotted line in  Fig.\ref{figdraco}. For comparison we also show
the  result from  the  analytical calculation  and  the same  quantity
relative to the smooth NFW halo of Draco. As in the case of the MW, we
note  that the  merger tree  approach gives  a comparable  yet smaller
result than the analytical one.

\cite{pbb,gpt08,Petal08}   showed   how   the   detectability   of   a
$\gamma$-ray  flux from such  a population  of sub-subhaloes  with the
Fermi-LAT telescope serendipitously depends on a very boosted particle
physics contribution.   The present and more accurate  results give an
even  lower  expected  flux,  thus  furtherly reducing  the  hope  for
detection,  unless an exotic  boost from  particle physics  enters the
play.   In  fact,  \cite{Baltz2008}  have  computed  the  map  of  the
Fermi-LAT sensitivity  to point sources of DM  annihilations, by using
the released Fermi-LAT response functions.   Draco lies in a region of
the sky where the $5\sigma$ detection  flux above 100 MeV in 1 year of
data taking is $\phi^{\rm 1  yr}_{5 \sigma} = 6 \times 10^{-9}$ \phcms
.  In order to obtain the afore-mentioned flux we have to multiply the
particle  physics  contribution by  the  highest  value of  $\Phi^{\rm
  cosmo}$ we  have obtained, that  is to say  along the line  of sight
pointing towards  the centre  of Draco.  Our  result is that  we would
need a  boost factor  of 600 (120)  when using the  already optimistic
scenarios  where $m_\chi  = 100$  (40)  GeV and  $\sigma v  = 3  \times
10^{-26}$ \cms.

In the case of Draco and  for the $B_{z_0,ref}$ model, we can conclude
that no higher order computations are necessary (sub-sub-subhaloes and
so  on).   In  fact,  we   have  computed  the  boost  factor  due  to
substructures  for any  mass  of  the parent  halo  from $10^{-3}$  to
$10^{10} \ M_\odot$,  that is to say the ratio of  the integral of the
density squared over  the whole galaxy including subhaloes  to the same
quantity computed in the case of  a smooth halo, which gives the boost
to the  $\gamma$-ray flux due  to substructures for  point-like halos.
We found that  its value is pretty flat and close  to 1, only slightly
larger for larger mass haloes, which however does not appear point-like
at the  distance of Draco. This  means that no boost  is obtained when
considering  sub-substructures  for point-like  subhaloes.   As far  as
massive  haloes  (with  mass   greater  than  $10^3  \  M_\odot$)  are
concerned, which are not point-like  at the distance of Draco, we have
to consider the fact that, as in  the case of Draco, the boost must be
distributed spatially. Indeed, this  spatial distribution of the boost
factor leads to  an even smaller value of the  expected flux along the
line  of sight  toward  the centre  of  the halo,  where  the flux  is
higher. This means that  including sub-substructures does not increase
the total $\gamma$-ray flux from the subhalo.

\section{Discussions and Conclusions}
\label{deconc}
In this paper we used the SL99 merger tree technique to study the mass
accretion  history  in  satellite  of  a Milky-Way  sized  halo.   The
partition code has been generalised for a $\mathrm{\Lambda CDM}$ power
spectrum,  considering   a  microsolar  mass  resolution   of  $m_J  =
10^{-6}\msun$.  The  MW-main progenitor  halo has been  followed along
consecutive  time steps from  the present  day down  to when  its mass
dropped below the resolution limit.  From the satellite haloes we have
identified a sample of Draco-like systems and also followed them along
their  merger tree, starting  from their  accretion redshift  into the
MW-main halo  progenitor. Both the  Milky-Way and the  Draco satellite
mass  function turned to  be equivalent,  testing the  universality of
this  distribution. The  partition technique  allowed us  to  build up
models for the present-day subhalo mass function for both MW and Draco
-- considering only haloes that they accreted along the main branch.

We then  computed the expected $\gamma$-ray flux  from DM annihilation
in such a  population of substructures.  We found  that the prediction
for the  $\gamma$-ray flux  is indeed more  pessimistic than  the ones
obtained in previous estimates {\bf which were not correctly taking into account the embedding of
small scale sub-subhaloes within subhaloes, and therefore adding their contribution to the 
host halo total flux instead of to the subhalo they belong to.} 
Detection with the Fermi-LAT
telescope is  probably out  of the discovery  range of  the satellite,
unless   some  exotic   particle  physics   could  boost   the  signal
significantly.

While  we  were  writing  this  paper, the  results  of  the  Aquarius
simulation came  out \citep{Springel}, predicting a  smaller number of
subhaloes  and a shallower  internal slope  for the  subhaloes density
profiles.  We  will not  repeat our analysis  in their model  since it
would  fourthly reduce  the expected  flux, but  the reader  should be
aware that  the results presented  in this paper,  though pessimistic,
are  upper limits  for the  expected  signal from  DM annihilation  in
galactic subhaloes.

\section*{acknowledgements}
CG thanks the partial support for  this work given by the Italian PRIN
and ASI.   Thanks also  to R.  K.   Sheth for very  useful discussions
during the time spent in Philadelphia at the end of $2007$.

\bibliographystyle{mn2e}

\label{lastpage}
\end{document}